\documentclass[sigconf]{acmart}

\def\BibTeX{{\rm B\kern-.05em{\sc i\kern-.025em b}\kern-.08emT\kern-.1667em\lower.7ex\hbox{E}\kern-.125emX}}
 
\usepackage{fancyhdr}
\usepackage{amsfonts, amsmath, amssymb}
\usepackage{booktabs}
\usepackage{multirow}
\usepackage{graphicx, color, setspace, url, verbatim, afterpage}
\usepackage{stfloats}
\usepackage{xspace}
\usepackage{paralist}
\usepackage{textcomp}
\usepackage{mathtools}
\usepackage{listings}
\usepackage{algorithm}
\usepackage[noend]{algpseudocode}
\usepackage{breqn}
\usepackage{subcaption}
\usepackage{enumitem}

\newcommand{\secref}[1]{\S~\ref{#1}}
\renewcommand{\eqref}[1]{Equation~\ref{#1}}
\newcommand{\secsref}[2]{\S\S~\ref{#1} and~\ref{#2}}
\newcommand{\figref}[1]{Figure~\ref{#1}}
\newcommand{\tabref}[1]{Table~\ref{#1}}
\newcommand{\minititle}[1]{\textbf{{#1}}}

\newcommand{\ignore}[1]{}

\def\ie{{\it i.e.}}
\def\eg{{\it e.g.}}

\def\doorbell{\emph{DoorBell}}
\def\llppost{\emph{LLP\_post}}
\def\llpprog{\emph{LLP\_prog}}
\def\hlppost{\emph{HLP\_post}}
\def\hlptxprog{\emph{HLP\_tx\_prog}}
\def\hlprxprog{\emph{HLP\_rx\_prog}}
\def\wire{\emph{Wire}}
\def\switch{\emph{Switch}}
\def\network{\emph{Network}}
\def\pcie{\emph{PCIe}}
\def\rctomem[#1]{\emph{RC-to-MEM(#1B)}}
\def\latency{\emph{Latency}}
\def\inj{\emph{Inj\_overhead}}
\def\msginj{\emph{Msg\_inj\_overhead}}
\def\cputime{\emph{CPU\_time}}
\def\post{\emph{Post}}
\def\txprog{\emph{Post\_prog}}
\def\misc{\emph{Misc}}

\copyrightyear{2019} 
\acmYear{2019} 
\setcopyright{acmlicensed}
\acmConference[ICPP 2019]{48th International Conference on Parallel Processing}{August 5--8, 2019}{Kyoto, Japan}
\acmBooktitle{48th International Conference on Parallel Processing (ICPP 2019), August 5--8, 2019, Kyoto, Japan}
\acmPrice{15.00}
\acmDOI{10.1145/3337821.3337910}
\acmISBN{978-1-4503-6295-5/19/08}

\begin{document}

%
\title{Breaking Band: A Breakdown of High-performance Communication}

%
\author{Rohit Zambre}
\email{rzambre@uci.edu}
\affiliation{%
	\institution{EECS, University of California, Irvine, USA}
}

\author{Megan Grodowitz}
\email{Megan.Grodowitz@arm.com}
\affiliation{%
	\institution{Arm Research, USA}
}

\author{Aparna Chandramowlishwaran}
\email{amowli@uci.edu}
\affiliation{%
	\institution{EECS, University of California, Irvine, USA}
}

\author{Pavel Shamis}
\email{Pavel.Shamis@arm.com}
\affiliation{%
	\institution{Arm Research, USA}
}

%
\renewcommand{\shortauthors}{Zambre et al.}

%
\begin{abstract}
	
	The critical path of internode communication on large-scale systems is composed of multiple components. When a supercomputing application initiates the transfer of a message using a high-level communication routine such as an \texttt{MPI_Send}, the payload of the message traverses multiple software stacks, the I/O subsystem on both the host and target nodes, and network components such as the switch. In this paper, we analyze where, why, and how much time is spent on the critical path of communication by modeling the overall injection overhead and end-to-end latency of a system. We focus our analysis on the performance of small messages since fine-grained communication is becoming increasingly important with the growing trend of an increasing number of cores per node. The analytical models present an accurate and detailed breakdown of time spent in internode communication. We validate the models on Arm ThunderX2-based servers connected with Mellanox InfiniBand. This is the first work of this kind on Arm. Alongside our breakdown, we describe the methodology to measure the time spent in each component so that readers with access to precise CPU timers and a PCIe analyzer can measure breakdowns on systems of their interest. Such a breakdown is crucial for software developers, system architects, and researchers to guide their optimization efforts. As researchers ourselves, we use the breakdown to simulate the impacts and discuss the likelihoods of a set of optimizations that target the bottlenecks in today's high-performance communication.
\end{abstract}

%
%


%

%

%
\maketitle

\section{Introduction}
\label{sec:intro}

\vspace{0.5em}
\emph{"To measure is to know."}\hfill--- Lord Kelvin
\vspace{0.5em}

\begin{figure}[tbp]
	\begin{center}
		\includegraphics[width=0.49\textwidth]{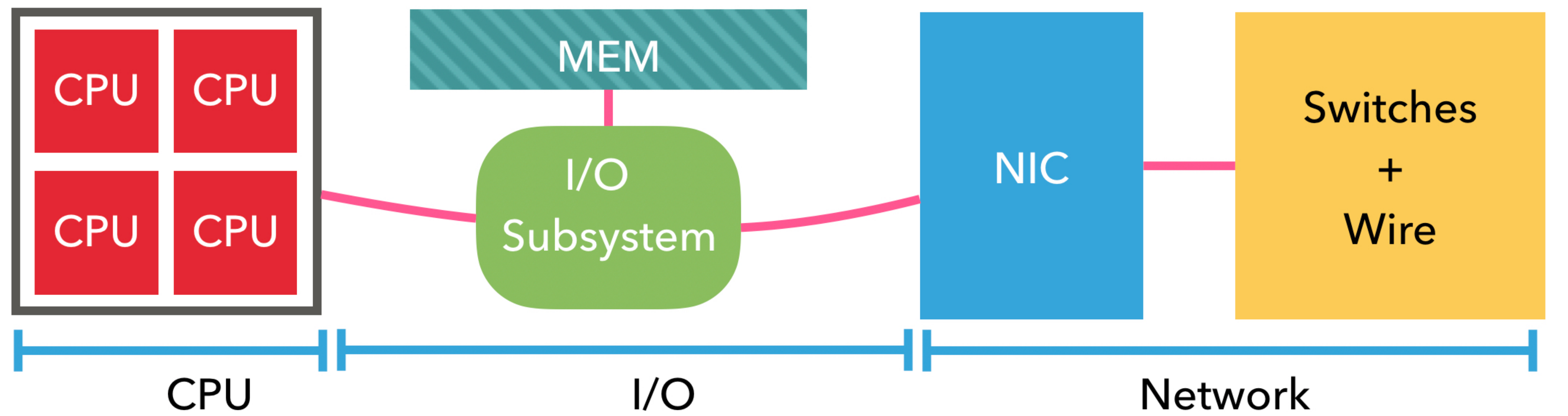}
	\end{center}
	\caption{Components involved in the transmission of a message (on one end).}
	\vspace{-1em}
	\label{fig:components}
	\vspace{-1em}
\end{figure}

Internode communication is the crux of supercomputing.
We can classify the various components involved in sending a message
into one of three categories: CPU, I/O, or network fabric,
as shown in \figref{fig:components}. Software stacks on the CPU
include the Message Passing Interface (MPI) and the communication
protocol processing in the underlying communication frameworks.
I/O encompasses subsystems on the processor chip such as PCI Express (PCIe).
Network components are the high-performance interconnect's switches and physical wire. Each of
these components on the critical path of communication poses an
opportunity for optimization. However, blindly optimizing all of the
components is impractical considering the technical challenges
associated with each and the wide variety of use cases. For example,
the latency of sending a large message is driven by the time spent
in the network components. Hence, optimizing the software stack for
this case would be a futile effort. On the other hand, the time spent
in the software stack during the propagation of a small message is a
considerable portion of the overall latency and, hence, optimizing the
time spent in the CPU would be beneficial. Therefore, it is
important to understand where to focus our optimization efforts. 

With the apparent end of Moore's law, the architectures of recent servers
are now featuring a large number of cores per node~\cite{tx2,xeon}, a trend
that is likely to continue moving forward~\cite{thakur2010mpi}. Furthermore,
other on-node resources such as memory, translation lookaside buffers, and
network-hardware registers are not growing at the same rate. Since developers
desire to solve the same problem faster on newer machines, they need to rely
on strong scalability with the decreasing amount of memory per core (assuming a
static split with a process per core). At the limits of strong scaling lies
fine-grained communication, that is, each core participates in communication,
eliminating the need to synchronize with the cores on a node. Since each core
communicates independently of the others, the size of the messages involved in
communication is small. Hence, we focus our analysis on the communication performance of small
messages since it is a critical factor in overall performance.

CPU, I/O, and network equally contribute to the
communication performance of small messages; the times spent in each of the categories are on
the same order of magnitude on state-of-the-art systems (we demonstrate
this in \secref{sec:complete}). Hence, optimizations of each category's constituents
would be beneficial. This raises the question: \emph{how much will optimizing
component X improve the overall communication performance?} The answer
to this question can guide the research and engineering efforts of software
developers, system architects, and the HPC community at large. Typically, one
measures the limits of a system's communication performance using
injection-rate and latency tests. But such measurements do not
inform the researcher where time is being spent or why the performance of one
version of the system varies from that of another.

In this paper, we answer the posed question by analyzing the
time spent in state-of-the-art software and system components during
the transmission of messages. We classify the components into
two levels: low and high. Low-level components include those
that are not exposed to a typical end-user of an HPC system. These include
the low-level communication framework (\eg\ Verbs), the I/O subsystem (\eg\ PCIe),
and network components (\eg\ Mellanox InfiniBand fiber). High-level components
include programming model frameworks such as MPI.

\minititle{Contributions and findings}. This paper makes the following
contributions.

\begin{enumerate}[leftmargin=*]
	\item \textbf{Detailed breakdown}. As a first step to answer the posed
	question, we construct analytical models of the overall injection overhead and
	end-to-end latency of a system. Our models explain where and why time is
	spent during the transmission of a message. By attributing times to the models'
	constituents using precise CPU timers and traces from a PCIe analyzer, we show
	how much time is spent in low-level (\secref{sec:lowlevel}) and high-level
	(\secref{sec:highlevel}) components, and thus present a detailed breakdown of
	high-performance communication. Our analytical models estimate the observed
	performance within a 5\% margin of error on Arm ThunderX2. This work is the
	first of its kind on Arm. We use the breakdown to provide key insights in
	\secref{sec:complete}. 

	\item \textbf{Measurement methodology}. We present a detailed methodology to
	measure the overhead of each component such as the PCIe wire, the interconnect's
	wire, etc. Researchers with access to a similar analysis infrastructure described in
	\secref{sec:setup} can then measure overheads for components of their interest
	using our methodology.
	
	\item \textbf{Simulated optimizations}. Finally, we answer the aforementioned
	question in \secref{sec:simulated} through a what-if analysis. We discuss the
	impact and likelihood of a set of optimizations that target the CPU, I/O, and
	network components of high-performance communication.

\end{enumerate}

 \section{Background}
\label{sec:background}

\begin{figure}[tbp]
	\begin{center}
		\includegraphics[width=0.5\textwidth]{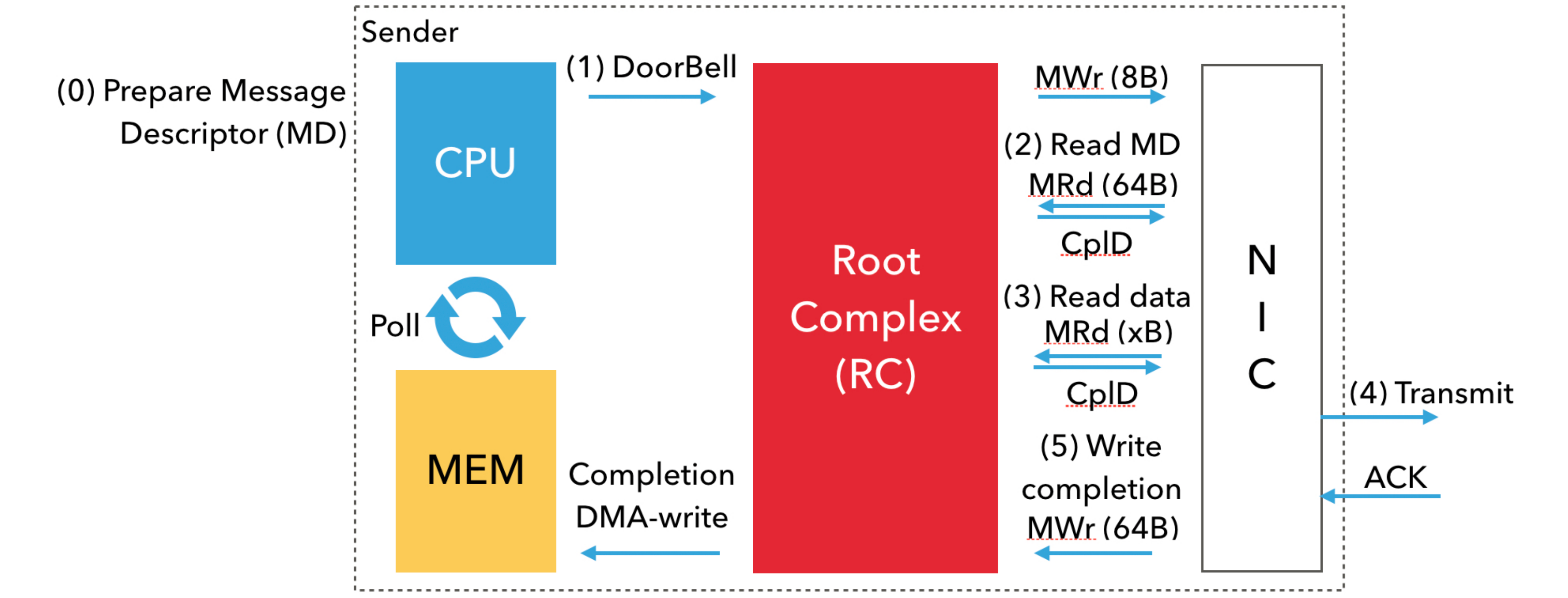}
	\end{center}
	\vspace{-1em}
	\caption{PCIe transactions and mechanisms on sender node to transmit
		data over wire.}
	\label{fig:ibmech}
	\vspace{-1em}
\end{figure}

The Network Interface Cards (NICs) of modern interconnects
are typically connected to the processor chip on the node 
as a PCI Express (PCIe) device. In this section, we
delineate the transmission and completion of messages in the
context of the PCIe fabric.

\minititle{PCI Express}. The main conductor of the PCIe subsystem is the Root Complex
(RC). It connects the processor and memory to the PCIe fabric.
The peripherals connected to the PCIe fabric are called PCIe
endpoints. The PCIe protocol consists of three layers: the 
Transaction layer, the Data Link layer, and the Physical layer.
The first, the upper-most layer, describes the type
of transaction occurring. In this paper, two types of
Transaction Layer Packets (TLPs) are relevant: Memory
Write (MWr), and Memory Read (MRd). Unlike the standalone MWr
TLP, the MRd TLP is coupled with a Completion with Data (CplD)
transaction from the target PCIe endpoint which contains the data
requested by the initiator.
The Data
Link layer ensures the successful execution of all transactions
using Data Link Layer Packet (DLLP) acknowledgements
(ACK/NACK) and a credit-based flow-control mechanism. An initiator
can issue a transaction as long as it has enough credits for that
transaction. Its credits are replenished when it receives Update Flow
Control (UpdateFC) DLLPs from its neighbors. Such a flow-control
mechanism allows the PCIe protocol to have multiple outstanding
transactions.

\minititle{Mechanisms of a high-performance interconnect}.
From a CPU programmer's perspective, there exists a \emph{transmit
queue} (TxQ) and a \emph{completion queue} (CQ). The user posts their
message descriptor (MD) to the transmit queue, after which they
poll on the CQ to confirm the completion of the
posted message. The user could also request to be notified with
an interrupt regarding the completion. However, the polling approach
is latency-oriented since there is no context switch to the kernel in
the critical path. The actual transmission of a message over the
network occurs through coordination between the processor chip
and the NIC using memory mapped I/O (MMIO) and direct memory
access (DMA) reads and writes. We describe these steps below using
\figref{fig:ibmech}.

\begin{enumerate}[leftmargin=*]
	\setcounter{enumi}{-1}
	
	\item The user first enqueues an MD into the TxQ.
	The network driver then prepares the device-specific MD that
	contains headers for the NIC, and a pointer to the payload.
	
	\item Using an 8-byte atomic write to a memory-mapped location,
	the CPU (the network driver) notifies the NIC that a message is ready
	to be sent. This is called \emph{ringing the} \doorbell. The RC executes
	the \doorbell\ using a MWr PCIe transaction.
	
	\item After the \doorbell\ ring, the NIC fetches the MD
	using a DMA read. A MRd PCIe transaction conducts the DMA read.
	
	\item The NIC will then fetch the payload from a registered memory
	region using another DMA read (another MRd TLP). Note
	that the virtual address has to be translated to its physical address
	before the NIC can perform DMA-reads.
	
	\item Once the NIC receives the payload, it transmits the
	read data over the network. Upon a successful transmission, the NIC
	receives an acknowledgment (ACK) from the target-NIC.
	
	\item Upon the reception of the ACK, the NIC will DMA-write (using
	a MWr TLP) a completion (64 bytes in Mellanox InfiniBand) to the CQ
	associated with the TxQ. The CPU will then poll for this
	completion to make progress.
\end{enumerate}

In summary, the critical data path of each post entails one MMIO write,
two DMA reads, and one DMA write. The DMA-reads translate to
round-trip PCIe latencies which are expensive.

A faster way to send a message that eliminates the PCIe round-trip 
latencies is \emph{Programmed I/O (PIO)}. With PIO, the CPU copies the MD
as a part of the \doorbell. Thus, the NIC doesn't need
to DMA-read the MD. Another feature for small payloads is
\emph{inlining} which means that the payload is a part of the MD.
Hence, when the NIC receives the MD, it does not need to
DMA-read the payload. Typically, communication frameworks, such as UCX, combine PIO
with inlining. This eliminates both the DMA-reads (steps (2) and (3)).
In Mellanox InfiniBand, the PIO occurs in 64-byte chunks. Note that the
CPU does more work in PIO (64-byte copy instead of an 8-byte write)
and inlining (\texttt{memcpy}). However, the increase in CPU's work
compared to the benefit gained from elimination of PCIe round-trip
latencies is minimal.

\section{Evaluation setup}
\label{sec:setup}

To measure the breakdown of time spent in components we use a
system of two nodes, node 1 and node 2, that are connected
to each other using a high-performance interconnect. Node 1 plays
the role of the initiator in our following experiments. We use the
CPU's timers to measure the time spent in software. To measure the
time spent in other components, we use traces from a PCIe analyzer.
Note that one can use this analysis infrastructure for any CPU or
interconnect of interest.

\begin{figure}[tbp]
	\begin{center}
		\includegraphics[width=0.49\textwidth]{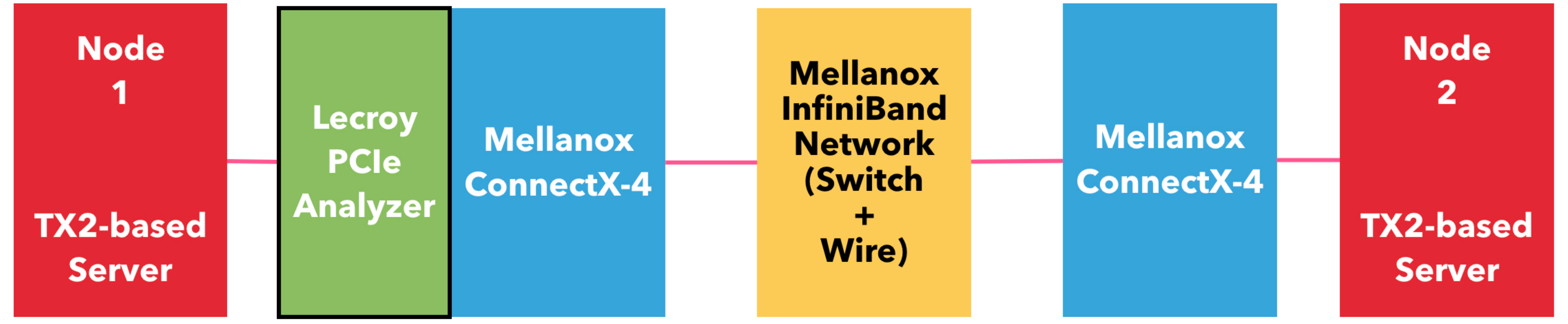}
	\end{center}
	\vspace{-1em}
	\caption{Two-node setup with PCIe analyzer on node 1.}
	\label{fig:pcietestbed}
	\vspace{-1.5em}
\end{figure}

We choose a state-of-the-art ThunderX2-based (TX2) server (running at 2 GHz)
for the nodes and TOP500-popular Mellanox InfiniBand~\cite{ibtop500} as the high-speed interconnect.
Specifically, we use ConnectX-4, a recent Mellanox InfiniBand adapter,
and attach it to the node through a PCIe slot. A Lecroy PCIe analyzer
sits just before the NIC on node 1, as shown in \figref{fig:pcietestbed}.
The overhead of the PCIe analyzer is negligible as we did not observe
any difference in performance with and without it. Larsen et al~\cite{larsen2015reevaluation}
observe the same. The analyzer is a passive instrument
that allows data to pass through fully unaltered~\cite{lecroy}.

For our software stack, we use the CH4 device of MPICH~\cite{raffenetti2017mpi}
with Unified Communication (UCX)~\cite{shamis2015ucx} as the
underlying communication framework. Specifically, we use UCX's
\emph{rc\_mlx5} transport which is UCX's implementation of the
data-path operations, such as posting to the transmit queue and
polling from the completion queue, for modern Mellanox InfiniBand
adapters.

To measure time spent in the CPU, we instrument relevant code
with UCX's UCS profiling infrastructure~\cite{ucs_profiling}, which
internally reads the \texttt{cntvct\_el0} register timer preceded by an \texttt{isb}
for \texttt{aarch64}. The mean overhead of this infrastructure is 49.69
nanoseconds (a standard deviation of 1.48 for 1000 samples); we report
software measurements in the rest of the paper after removing this overhead.

Each reported CPU or PCIe analyzer measurement is a mean of at least
100 samples. While measuring time of a component, we do not
simultaneously measure time in any other component to minimize any
effects of artificial slowdowns caused by the timer infrastructure. Hence, we do not require
synchronized timers.

\section{Breakdown of the Lower Level}
\label{sec:lowlevel}

In this section, we present a detailed breakdown of time spent in the
low-level components. These include the low-level 
communication protocol (LLP),  the I/O subsystem, and network
components. The LLP software drives the I/O and network hardware.
We first define terminology for time spent in each of the low-level
components.


\begin{itemize}[leftmargin=*]
	\item \llppost\ -- LLP performing a PIO
	post of one 8-byte message.
	\item \llpprog\ -- LLP dequeuing one entry of
	the completion queue during the progress of an operation.
	\item \pcie\ -- payload traversing PCIe between RC and NIC.
	\item \wire\ -- payload traversing the physical wire of the interconnect.
	\item \switch\ -- overhead added by a network switch.
	\item \network\ -- the total time in the interconnect (\wire\ + \switch).
	\item \rctomem[x] -- RC writing an x-byte payload to memory.
\end{itemize}

We use UCX's low-level transport API, UC-Transports (UCT) for our
LLP driver. It abstracts the capabilities of the  various hardware architectures with minimal software overhead. The UCT driver runs
the UCX \texttt{perftest}'s injection-rate and ping-pong style latency
microbenchmarks, namely the \texttt{put\_bw},
and \texttt{am\_lat} tests, with a single thread. The \texttt{put}
test corresponds to RDMA-writes while the \texttt{am}\footnote{\texttt{am}
	is short for active messages, terminology that describes
	send-receive style messaging in UCX.} test corresponds to
send-receive semantics. Each message is 8 bytes, the size of
a \texttt{double}.

\subsection{Breakdown of the LLP}
\label{sec:llpbreakdown}

The LLP implements the HW/SW interface required to transmit
a message and confirm its completion. The network driver
(software) invokes the NIC (hardware) directly after correctly
preparing resources and registers needed by the NIC during an
\llppost. The following details the steps involved in an \llppost.

\begin{enumerate}[leftmargin=*]
	\item Prepare MD -- this involves the time taken to
	write the control segment of the descriptor. It also involves a \texttt{memcpy} of the
	small payload when inlining is used.
	\item A store memory barrier -- this ensures that the MD
	is completely written before the CPU signals the NIC. This barrier is 
	relevant only for a weak memory model (\texttt{dmb st} on \texttt{aarch64}).
	\item \doorbell\ counter increment -- the NIC reads a
	\doorbell\ counter to perform speculative reads. The CPU updates this
	counter before writing to the NIC.
	\item A store memory barrier -- this ensures that the NIC sees the update
	to the \doorbell\ counter before any subsequent write to its device memory.
	\item PIO copy -- this is the CPU's write to the memory-mapped device memory
	instructing the NIC to transmit the message. Device memory is
	typically an uncached, buffered memory region that supports out-of-order
	writes. For the TX2-based server in our setup, we use Device-GRE memory
	for the memory-mapped location. Though there would be a store memory
	barrier (\texttt{dsb st}) after the PIO copy to flush the data to the NIC, we
	observed experientially that this flush is not necessary for the
	microarchitecture of the TX2-based server. The PIO copy
	of an 8-byte message is one 64-byte chunk in Mellanox InfiniBand (see \secref{sec:background}).
\end{enumerate}

Similarly, the LLP reads the designated memory location (where
the NIC DMA-writes its completions) during an \llpprog, the
progress of an operation. This progress operation constitutes a
load memory barrier for \texttt{aarch64}'s weak memory model to ensure that
the read for a completion queue entry occurs before subsequent
updates to data structures.

\begin{figure}[tbp]
	\begin{center}
		\includegraphics[width=0.49\textwidth]{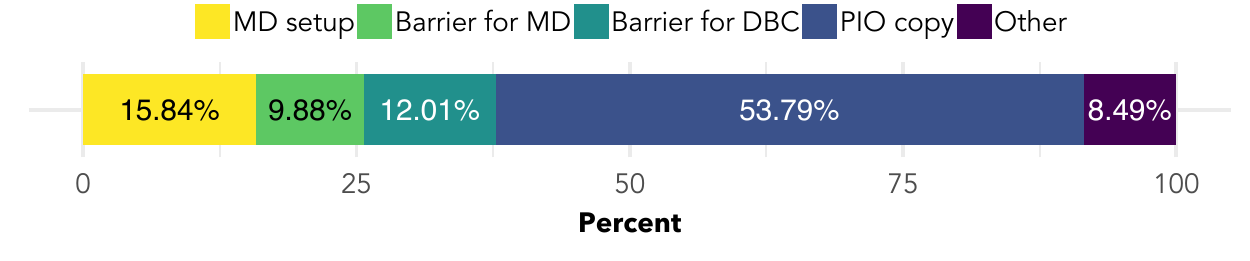}
	\end{center}
	\vspace{-1em}
	\caption{Breakdown of time in an \llppost\ (MD: message descriptor;
		DBC: \doorbell\ counter).}
	\label{fig:llppostbreakdown}
	\vspace{-1.5em}
\end{figure}

\minititle{Measuring LLP and its breakdown.} We measure \llppost\ and
\llpprog\ by wrapping the UCS profiling infrastructure around the calls
to \texttt{uct\_ep\_put\_short} and \texttt{uct\_worker\_progress}.
We use the same technique around the relevant regions of code in
the implementation of \texttt{uct\_ep\_put\_short} to measure the time
in each of the categories of an \llppost. While these categories are critical
components of an \llppost, they do not account for other miscellaneous time
such as the function call overhead, branches to decide code path, etc. We
compute this time by taking the difference of \llppost\ and
the sum of the times spent in the categories. \tabref{tab:measuredtimes}
reports the times for \llppost, \llpprog, and each category of \llppost.
\figref{fig:llppostbreakdown} shows the breakdown
of \llppost. Since the \llpprog\ contains only one critical category (the load
memory barrier), we don't show its breakdown. 

\subsection{Injection overhead}
\label{sec:injection}

Injection is the insertion of a message into the network.
The message is injected when the payload reaches the NIC.
We study the case when the user is transmitting messages
continuously since this represents a system's injection
limit. Then, the system's injection overhead,
\inj, is the time difference between messages arriving
at the NIC. This \inj\ explains why all the messages
in a burst do not reach the NIC at time zero. We first model the injection
overhead of PIO posts for a small message, then measure the overhead
according to the model, and finally validate it.

\minititle{Modeling injection overhead.} Since the depth of the
transmit queue (TxQ) is finite, the user cannot post indefinitely.
Polling the completion queue (CQ) serves as the dequeue semantic
for the TxQ. Hence, the user must poll in between
their posts to inject messages into the NIC. Say, the user polls
after every $p$ posts. If $p=1$, the depth of the TxQ is
not utilized and the post translates to a synchronous post, that
is, the user will be able to post the next message only after
the previous message has reached the target node (since the
completion is generated only when the host NIC receives an ACK
from the target NIC (see \secref{sec:background})).

To remove the overhead of waiting for a previous message
to complete, the user must choose a value of $p$ such
that the completion for an earlier message is available during
a poll. Such a value of $p$ depends on the value of \llppost\ and
the time taken to generate a completion, $gen\_completion$.
From \secref{sec:background}, we can deduce that
\begin{dmath*}
gen\_completion = 2 \times (\pcie + \network) + \rctomem[64]
\end{dmath*}
since the PCIe wire and the interconnect's network fabric are
traversed twice: first while transmitting the message to the target
NIC, and second while receiving the ACK from the target NIC
and writing the corresponding completion. A completion in
InfiniBand is 64 bytes and hence, the RC conducts a 64-byte
write to memory on behalf of the host NIC. Then, to remove
the overhead of waiting for a previous message, the lower
bound on $p$ is
\begin{dmath*}
p \ge gen\_completion / \llppost
\end{dmath*}
In our modeling
of the injection overhead, we assume that the user meets this 
lower bound on $p$.

Typically, the API of the network driver allows the user to
poll a batch of completions, reducing the overhead of 
expensive memory barriers and function calls~\cite{kalia2016design}.
Say the user polls $b$
number of completions in each batch. This means that the
user can post only $b$ posts in the next round of posts since
only $b$ entries have been dequeued from the TxQ.
Note that $b$ meets the lower bound mentioned
above. Additionally, the user could perform some miscellaneous
operations during the window of $b$ posts or $b$ polls. Let
$tot\_misc$ demarcate the cumulative time spent in these
other operations. Then, the overhead of the CPU to post a
message is
\begin{dmath*}
\cputime = \frac{b\times\llppost+ b\times\llpprog + tot\_misc}{b} = \llppost + \llpprog + \misc
\label{eq:cputime}
\end{dmath*}
where $\misc = tot\_misc/b$ is the miscellaneous
overhead amortized for each message.

Hence, on average, messages arrive at the RC every \cputime.
Since PCIe supports multiple outstanding requests
(see \secref{sec:background}), the RC initiates MWr PCIe
transactions targeting the NIC as soon as it receives
messages from the CPU. Considering that the RC is
implemented with hardware logic, the time it takes to
generate a transaction would be in the order of a few cycles.
Hence, we ignore its contribution to the injection overhead.
Note that the RC can generate transactions only if it has
enough credits. Otherwise, it needs to wait for
an UpdateFC DLLP from the NIC which would incur the
overhead of the PCIe wire between the NIC
and the RC (\pcie). Experientially, we observe that a single
core does not exhaust the credits for MWr transactions.
Hence, we do not model for the overheads imposed with
exhausted credits in this paper.

Once the message leaves the RC, it incurs \pcie\ before arriving
at the NIC. Hence, the injection overhead of a \emph{single}
message is
\begin{dmath*}
	\msginj = \cputime + \pcie
\end{dmath*}

\begin{figure}[tbp]
	\begin{center}
		\includegraphics[width=0.49\textwidth]{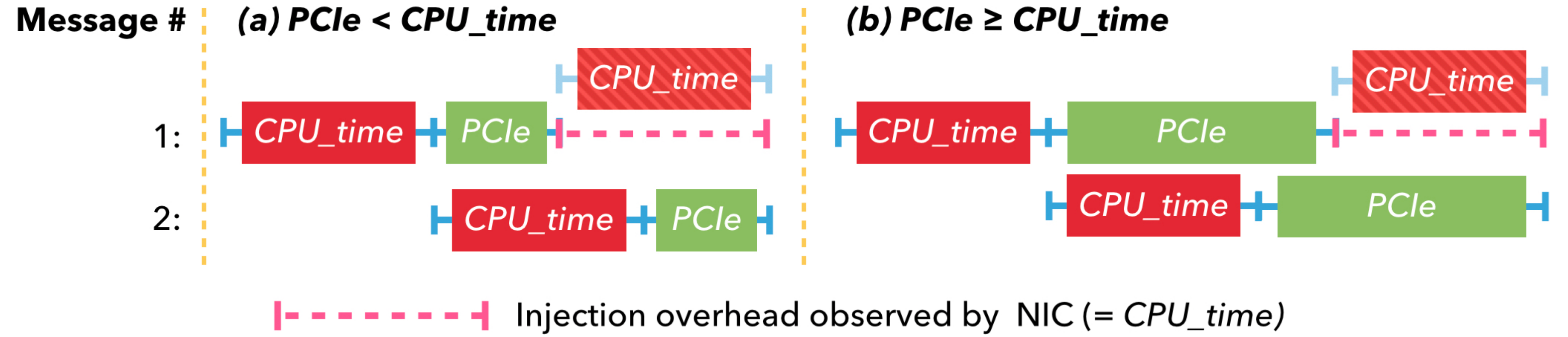}
	\end{center}
	\vspace{-1em}
	\caption{Injection overhead observed by the NIC.}
	\label{fig:nicinjection}
	\vspace{-1.5em}
\end{figure}

While \msginj\ describes the time taken by each message to
reach the NIC, it is not the same as the injection overhead
observed by the NIC, \inj, as we shall see next. When the system is
issuing messages continuously, the \cputime\ of the next
message overlaps with the \pcie\ of the previous one
(see \figref{fig:nicinjection}). Hence, the time difference between
the initiation of messages is \cputime. This holds true for any
relation of \pcie\ with \cputime\ (assuming that \pcie\ is not long enough to
exhaust the RC's credits). When $\pcie > \cputime$,
\pcie\ of the next message can also overlap with \pcie\ of the
previous one. Hence, from the perspective of the
NIC, the time difference between the arrival of messages is the
same as that between the initiation of messages, that is,
\begin{dmath}
	\inj = \cputime = \llppost + \llpprog + \misc
	\label{eq:llinjection}
\end{dmath}

Next, we measure the constituents of \inj. In \secref{sec:llpbreakdown},
we reported the times measured for \llppost\ and \llpprog. To
account for \misc, we first explain what occurs between
consecutive posts in UCX's \texttt{put\_bw} benchmark.

Every message in the benchmark generates a completion.
However, the benchmark polls for one completion every
16 posts. Hence, eventually the finite depth of the TxQ is fully
utilized after which an \llppost\ results in a "busy" post, that is,
an \llppost\ fails since an \llpprog\ must occur before the next
successful \llppost. Thus, in the average case, after every
successful \llppost, there occurs a busy post. Additionally, the
benchmark records a timestamp and updates its injection-rate
measurements after every \llppost. \tabref{tab:measuredtimes}
reports the times for a "Busy post" and a "Measurement update"
measured using the UCS profiling infrastructure wrapped around
the relevant code paths; $\misc = 56.58$ nanoseconds.

\begin{table}
	\centering
	\caption{Measured times of various components.}
	\label{tab:measuredtimes}
	\vspace{-1em}
	\begin{tabular}{  | r | l |}
		\hline
		\textbf{Component} & \textbf{Time (ns)} \\ \hline
		Message descriptor setup & 27.78 \\ \hline
		Barrier for message descriptor & 17.33 \\ \hline
		Barrier for \doorbell\ counter & 21.07 \\ \hline
		PIO copy (64 bytes) & 94.25 \\ \hline
		Miscellaneous in \llppost & 14.99 \\ \hline
		\llppost\ (total of above) & 175.42 \\ \hline \hline
		\llpprog & 61.63 \\ \hline \hline
		Busy post & 8.99 \\ \hline
		Measurement update & 49.69 \\ \hline
		\misc\ in $Inj\_overhead$ (total of above) & 58.68 \\ \hline \hline
		\pcie\ for a 64-byte payload & 137.49 \\ \hline \hline
		\wire\ & 274.81 \\ \hline
		\switch\ & 108 \\ \hline
		\network\ (total of above) & 382.81 \\ \hline \hline
		\rctomem[8] & 240.96 \\ \hline \hline
		\texttt{MPI\_Isend} in MPICH  & 24.37 \\ \hline \hline
		\texttt{MPI\_Isend} in UCP  & 2.19 \\ \hline \hline
		Callback for a completed \texttt{MPI\_Irecv} in MPICH & 47.99 \\ \hline
		Successful \texttt{MPI\_Wait} for \texttt{MPI\_Irecv} in MPICH & 293.29 \\ \hline \hline
		Callback for a completed \texttt{MPI\_Irecv} in UCP & 139.78 \\ \hline
		Successful \texttt{MPI\_Wait} for \texttt{MPI\_Irecv} in UCP & 150.51 \\ \hline
		
	\end{tabular}
	\vspace{-1.5em}
\end{table}

\begin{figure*}[tbp]
	\begin{center}
		\includegraphics[width=0.99\textwidth]{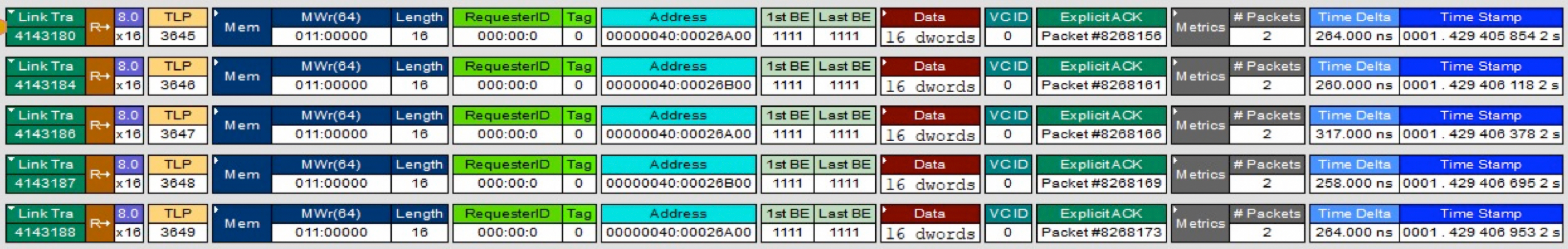}
	\end{center}
\vspace{-1em}
	\caption{PCIe trace of downstream PCIe transactions for UCX's RDMA-write injection-rate benchmark (\texttt{put\_bw}).}
	\label{fig:pcietrace}
	\vspace{-1em}
\end{figure*}

\begin{figure}[tbp]
	\begin{center}
		\includegraphics[width=0.49\textwidth]{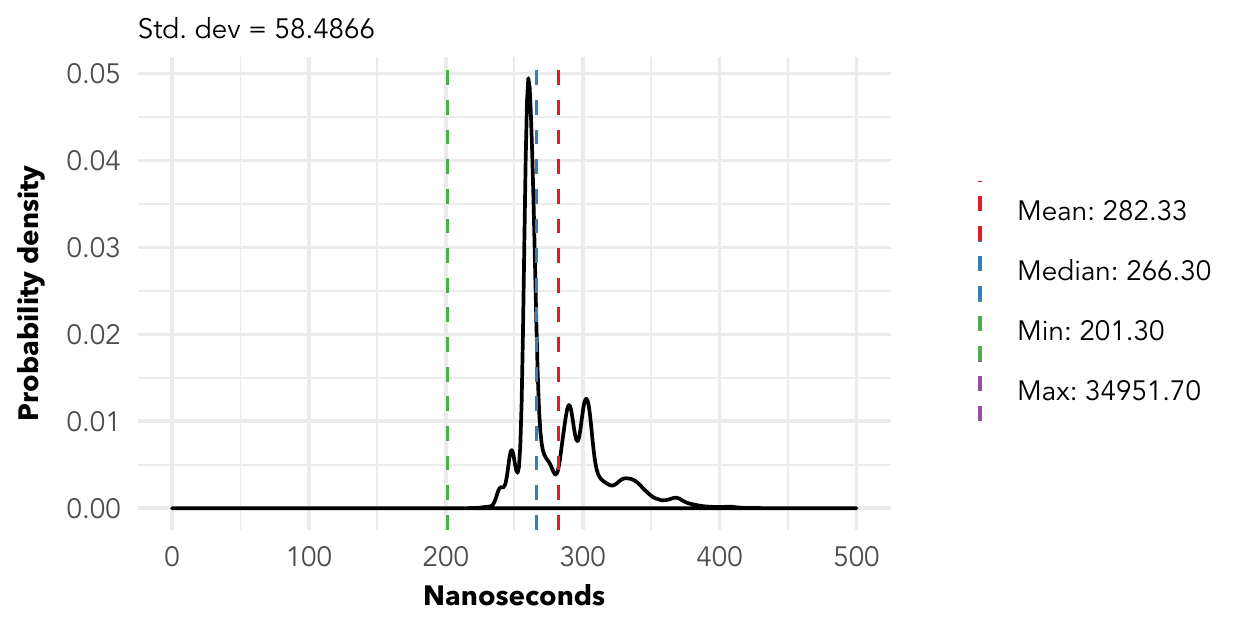}
	\end{center}
    \vspace{-1em}
	\caption[Injection overhead distribution]{Distribution\footnotemark of the observed injection overhead.}
	\label{fig:observedir}
\end{figure}

\begin{figure}[tbp]
	\begin{center}
		\includegraphics[width=0.49\textwidth]{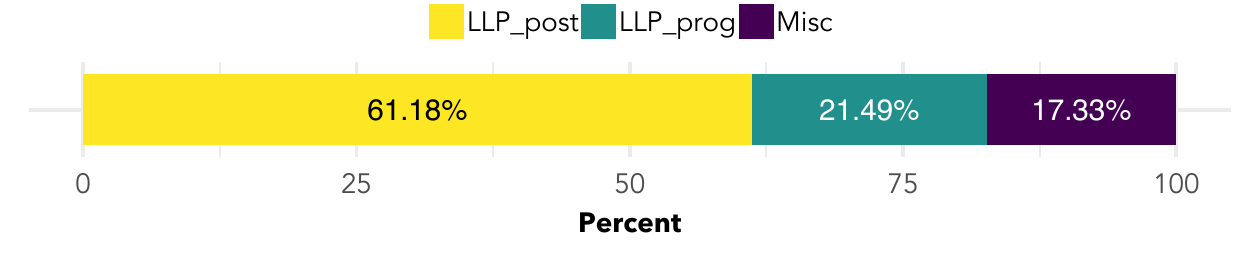}
	\end{center}
    \vspace{-1em}
	\caption{Breakdown of injection overhead with the LLP.}
	\label{fig:injection_breakdown}
	\vspace{-1em}
\end{figure}

\begin{sloppypar}
\minititle{Breakdown of injection overhead.} The PCIe trace of the
\texttt{put\_bw} test shows the observed injection overhead of
the system. \figref{fig:pcietrace} shows a snippet of the
PCIe trace after filtering for downstream (RC to NIC)
transactions. The data in each downstream transaction is 64 bytes
corresponding to the PIO post of an 8-byte payload. Every transaction
is associated with a timestamp. This timestamp corresponds to the time
when the PCIe analyzer observes the transaction. Since the PCIe analyzer
is sitting just before the NIC, these timestamps correspond to the times
at which the messages reach the NIC. Hence, calculating the delta
of the timestamp of consecutive transactions would result in the observed
\inj. \figref{fig:observedir} shows the distribution of this
overhead. The modeled injection overhead of \textbf{295.73} nanoseconds is within \textbf{5\%}
of \textbf{282.33} nanoseconds, the mean observed injection overhead. \figref{fig:injection_breakdown}
shows a percentage breakdown of \inj.
\end{sloppypar}

\subsection{Latency}
\label{sec:latency}

Latency is the total time incurred by a message starting from
the time the host node initiates the transfer to the time of writing
the payload in the destination buffer on the target node.

\minititle{Modeling latency.} We study the latency of a short
message transmitted using send-receive semantics. The
initiation of the transmission begins with an \llppost, after which
the message traverses the PCIe fabric and reaches the NIC.
The NIC then transmits the message over the network fabric
to reach the target node. On the target node, the NIC performs
a MWr PCIe transaction, which traverses the PCIe wire and 
instructs the RC to write the payload into the target node's
memory. Meanwhile the CPU on the target node has been
polling for its posted receive to complete. The user can use its
receive buffer only after a successful poll. Thus, for a payload
of size, \emph{x}, the time for latency is derived as follows,
\begin{dmath*}
\latency = \llppost + 2(\pcie) + \network + \rctomem[x] + \llpprog
\end{dmath*}

\begin{sloppypar}
Now, we measure the individual components that contribute to \latency. The value of an
\llppost\ is the same as the one measured in \secref{sec:injection},
that is, 175.42 nanoseconds.
\end{sloppypar}

\minititle{Measuring \pcie.} To measure \pcie, we first measure the
round-trip latency of the PCIe wire between the NIC and the RC.
Since the PCIe analyzer sits just before the NIC, any transaction
initiated by the NIC and the corresponding ACK DLLP from the RC
will give us the start and end time of the required round-trip.
For this purpose, we use the MWr transactions initiated
by the NIC during the DMA-write of completions. The
timestamp in the MWr transaction is the start time of the round
trip and that in the corresponding ACK DLLP is the end time.
Dividing this round-trip value by two is \pcie\ (the size of this MWr
transaction is the same as that of the PIO copy: 64 bytes).
We measure \pcie\ to be 137.49 nanoseconds.

\minititle{Measuring \network.} One way to measure \network\ would be to
first measure the time difference between when a PIO post reaches
the NIC and when the NIC receives an ACK from the target node
for that PIO post. Then, dividing that difference by two would 
correspond to \network\ since the difference entails a round-trip
\network\ latency. The timestamps on the PCIe trace of the
ping-pong style \texttt{am\_lat} benchmark allow us to employ this method. A
downstream 64-byte PCIe transaction corresponds to a ping
and the next upstream 64-byte PCIe transaction corresponds
to the ping's completion which is generated upon reception of
the ACK. Doing so, we measured the value of \wire\ to be 274.81 nanoseconds
for a direct NIC-to-NIC connection. If the NICs are connected via
a switch, the overhead of \switch\ is 108 nanoseconds. We measured
this by taking the difference between two latency
measurements: one with a switch involved and one without.

\footnotetext{Max is not shown in the figure due to the large value.}

\begin{figure}[tbp]
	\begin{center}
		\includegraphics[width=0.49\textwidth]{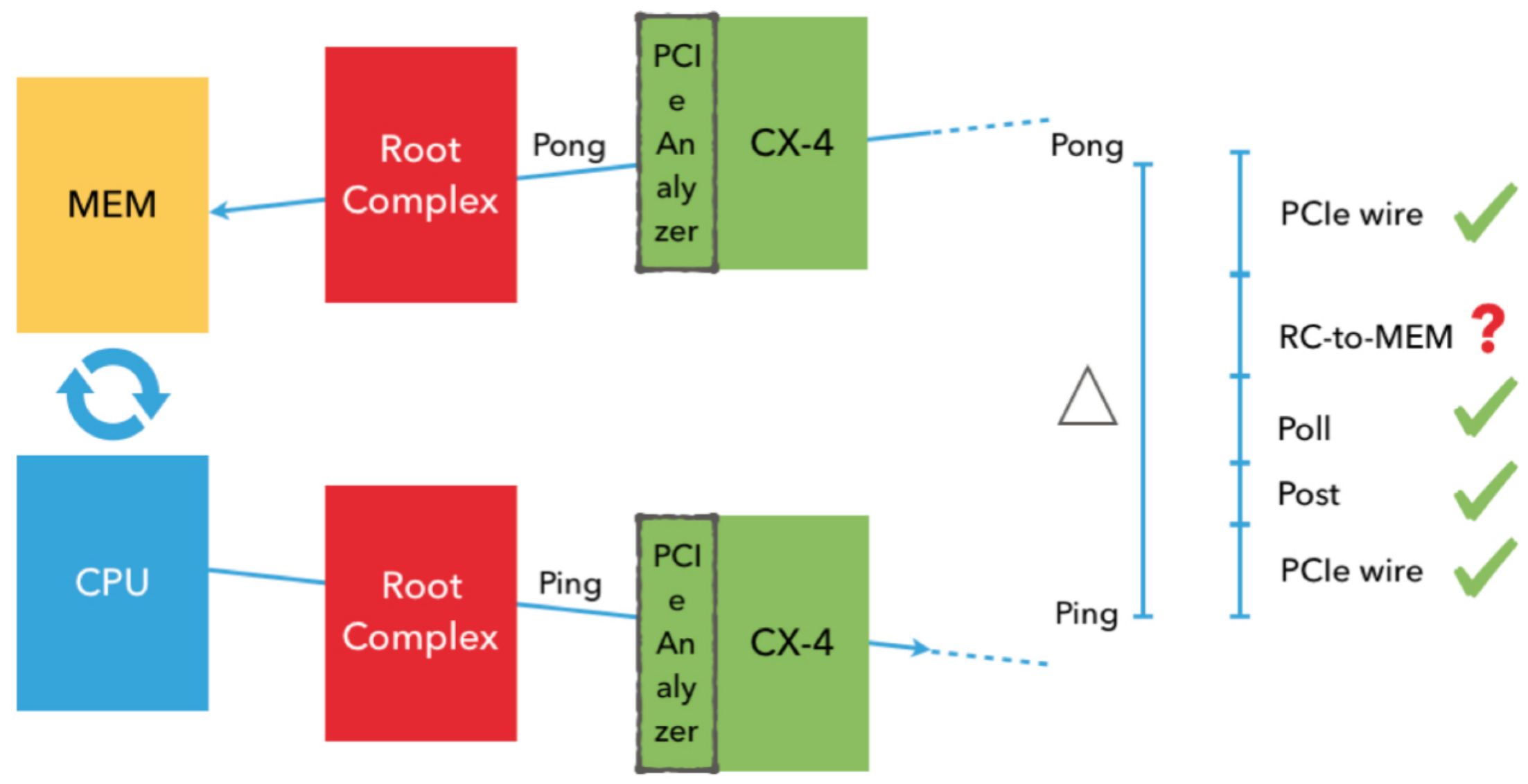}
	\end{center}
\vspace{-1em}
	\caption{Measuring \rctomem[x] using the time delta between an inbound
		pong and outbound ping on node 1.}
	\vspace{-1em}
	\label{fig:rctomemcalc}
\end{figure}

\minititle{Measuring \rctomem[8].} To measure \rctomem[8], we utilize
the timestamps on the PCIe trace data of the \texttt{am\_lat}
ping-pong benchmark. As shown in \figref{fig:rctomemcalc}, the
time difference between an incoming pong and outgoing ping
entails an \rctomem[8], two \pcie s (one for the inbound
pong and the other for the outbound ping), a \llpprog\ (successful poll),
and a \llppost\ (the ping). Once we measure the pong-ping difference
from the PCIe trace, we can compute the value of \rctomem[8] since
we have measured the values of the other components. This way, we
measured the value of \rctomem[8] to be 240.96 nanoseconds.

Plugging in our measured values (reported in \tabref{tab:measuredtimes})
into the latency model of a short message transmitted with send-receive
semantics, we have $\latency = \textbf{1135.8}$ nanoseconds.

\begin{figure}[tbp]
	\begin{center}
		\includegraphics[width=0.49\textwidth]{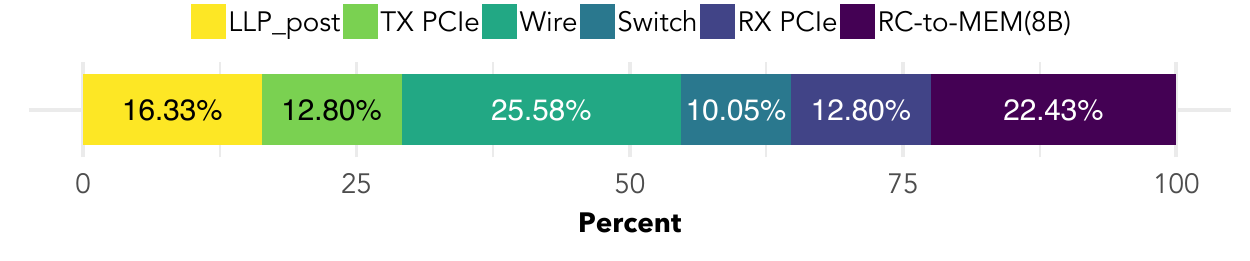}
	\end{center}
	\vspace{-1em}
	\caption{Breakdown of latency with the LLP.}
	\vspace{-1em}
	\label{fig:latencybreakdown}
\end{figure}

\begin{sloppypar}
\minititle{Breakdown of latency.} The observed latency from
UCX's \texttt{am\_lat} test is 1215 nanoseconds. The benchmark
measures a round-trip latency and then divides the
measurement by two to report the latency. Since a
measurement update occurs before the target responds with a pong,
we need to deduct half of "Measurement update" from \tabref{tab:measuredtimes} 
from the observed latency, which results in \textbf{1190.25} nanoseconds. The
modeled latency is within \textbf{5\% }of this observed latency. \figref{fig:latencybreakdown}
shows a percentage breakdown of latency.
\end{sloppypar}

\section{Breakdown of the Higher Level}
\label{sec:highlevel}

In this section, we present a breakdown of the time
spent in the high-level components of high-performance
communication. This comprises of the high-level
communication protocols (HLP). The most commonly
used programming model for large-scale parallel
systems today is MPI~\cite{thakur2010mpi}. Hence, at
the highest level of the software stack sits an MPI
library that implements the MPI standard. Modern
implementations, such as the CH4 device of MPICH,
rely on abstract communication frameworks, such
as UCX, so that the MPI libraries do not need to maintain
separate critical paths for all interconnects.

UCX in turn is composed of multiple components such
as UC-Transports (UCT) and UC-Protocols (UCP). UCT
is the LLP that we analyze in \secref{sec:llpbreakdown}.
UCP implements high-level communication protocols
such as collectives, message fragmentation, etc. using
the low transport-level capabilities exposed through
UCT. MPI libraries then use UCP to implement the
specifications of the MPI standard.

\begin{sloppypar}
We present a breakdown of time spent in
the HLP for a communication-initiation operation such as
\texttt{MPI\_Isend}, and a communication-progress operation
such as a successful (\ie\ no busy waiting) \texttt{MPI\_Wait}
corresponding to an \texttt{MPI\_Irecv}. 
\end{sloppypar}

\begin{figure}[tbp]
	\begin{center}
		\includegraphics[width=0.49\textwidth]{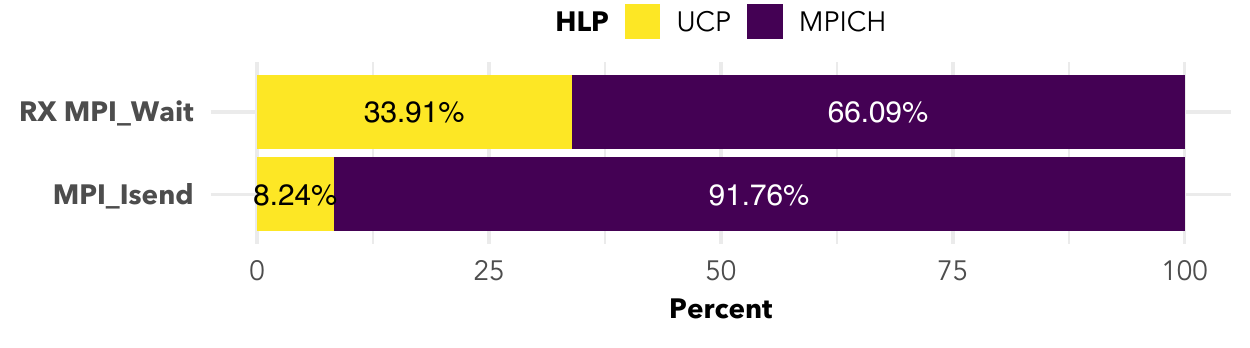}
	\end{center}
\vspace{-1.5em}
	\caption{Breakdown of time in HLP.}
	\vspace{-2em}
	\label{fig:mpichucpbreakdown}
\end{figure}

\begin{sloppypar}
\textbf{Measuring HLP and its breakdown.} In an
\texttt{MPI\_Isend}, the MPI library first decides how to best execute the
operation by checking if the data is contiguous, computing
which communication interface to use, etc. Ultimately it will call into
the UCP layer (\texttt{ucp\_tag\_send\_nb}) which will eventually execute
the LLP in the UCT layer (\texttt{uct\_ep\_am\_short}). To
measure the time spent in MPICH and UCP for an
\texttt{MPI\_Isend}, we first measure the total time of \texttt{MPI\_Isend},
the total time of \texttt{ucp\_tag\_send\_nb} inside MPICH, and the total
time of \texttt{uct\_ep\_am\_short} inside UCP by wrapping them with
the UCS profiling infrastructure. We can then measure the time spent
in MPICH and UCP by taking the differences of times between the
upper and lower layers. For example, subtracting the total time of
\texttt{ucp\_tag\_send\_nb} from that of \texttt{MPI\_Isend} gives us
the time spent in MPICH.
\end{sloppypar}

\begin{sloppypar}
Similarly, in an \texttt{MPI\_Wait}, the MPI library executes its progress
engine which ultimately calls into the UCP layer (\texttt{ucp\_worker\_progress}).
UCP will then ensure progress on all outstanding operations that have
posted by progressing the low-level UCT layer (\texttt{uct\_worker\_progress}).
When an operation completes, UCT executes a registered callback into the
upper UCP layer to update data structures that indicate the completion
of the operation. Similarly, the UCP callback also executes a registered callback
into the upper MPICH layer to indicate that the operation has completed. Note
that these callbacks are executed before returning from \texttt{uct\_worker\_progress}.
To measure the time spent in MPICH and UCP for an
\texttt{MPI\_Wait}, we measure the times spent in the registered MPICH and UCP
callbacks in addition to measuring the total times of \texttt{MPI\_Wait},
\texttt{ucp\_worker\_progress}, and \texttt{uct\_worker\_progress}. Since
the UCP callback entails the MPICH callback, we can measure the time
spent in the UCP callback alone by taking the difference between the total
times spent in the callbacks. We can then measure the time spent in MPICH
and UCP by taking the differences of times between the upper and lower layers
and adding in the time for the upper layer's registered callback. For example,
subtracting the total time of \texttt{ucp\_worker\_progress} from that of
\texttt{MPI\_Wait} and adding in the time of the MPICH callback gives us the time
spent in MPICH.
\end{sloppypar}

\tabref{tab:measuredtimes} reports the time spent in MPICH and UCP on top
of the LLP's HW/SW interface for an \texttt{MPI\_Isend} (26.56 nanoseconds in total),
and a successful \texttt{MPI\_Wait} for an \texttt{MPI\_Irecv} (443.8 nanoseconds in total).
\figref{fig:mpichucpbreakdown} shows their percentage breakdown.

\section{The Complete Picture}
\label{sec:complete}

In this section, we first present a breakdown of the overall injection
overhead and end-to-end latency including all the software, I/O,
and network components for send-receive communication. Then,
analyzing the breakdown, we note a number of insightful findings.

\begin{figure}[tbp]
	\begin{center}
		\includegraphics[width=0.49\textwidth]{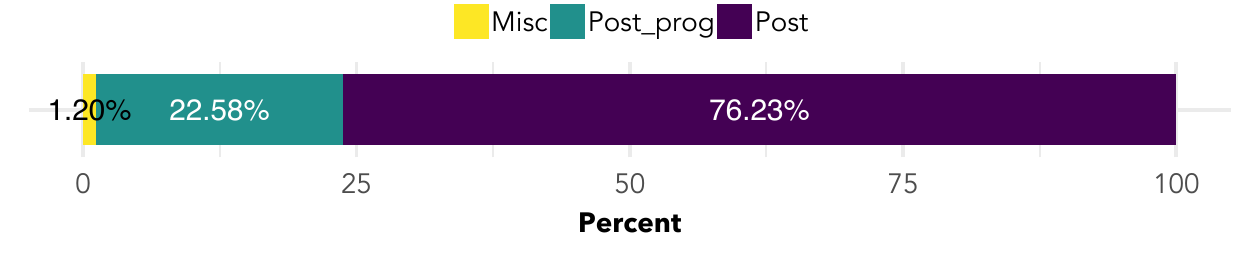}
	\end{center}
\vspace{-1em}
	\caption{Breakdown of the overall injection overhead.}
	\vspace{-1em}
	\label{fig:compinjbreakdown}
\end{figure}

\minititle{Overall injection overhead}. \secref{sec:injection} shows that
the injection overhead observed by the NIC for a single core is
governed by the rate at which the CPU can send messages to
the RC. \eqref{eq:llinjection} defines this injection overhead.
\cputime\ in \secref{sec:injection} involved only the overhead
of the LLP; in this section, we add in the overheads of the HLP to
complete the picture. We redefine \cputime\ as follows.
\begin{dmath}
	\cputime = \post + \txprog + \misc
	\label{eq:compinjection}
\end{dmath}
where \post\ is the total time taken by the HLP and LLP to initiate
an operation, and \txprog\ is the total overhead imposed by
both the HLP and LLP for the progress of a send-operation.
\post\ is the sum of \llppost\ and \hlppost, the time spent in the
HLP during the initiation of a message. For our setup, \hlppost\
equals 26.56 nanoseconds, which implies \post\ equals 201.98
nanoseconds. Before attributing times to \txprog\ and \misc,
we delineate certain caveats.

First, UCP schedules the successful execution of \llppost\ for busy
posts (see \secref{sec:injection}) during the progress of operations. Second, progress
for a bunch of initiated operations is typically
conducted with a batch-progress operation in the HLP such as
\texttt{MPI\_Waitall}. MPICH executes its progress engine until
all the operations listed in \texttt{MPI\_Waitall} complete. More
important, UCP reduces the overhead of progress using
unsignaled completions~\cite{kalia2016design}, which means the
NIC DMA-writes a completion only every $c$ operations to
indicate the completion of all $c$ operations ($c = 64$ in UCX).
Hence, the overhead of progress is amortized
over $c$ operations.

The first caveat implies that the progress of some operations
includes the overhead of initiation in the LLP. Since we 
already account for the successful posts of
busy posts in \post, we deduct the cumulative \llppost s
corresponding to the busy posts from the total time of
\texttt{MPI\_Waitall} for analytical purposes. We do so by keeping track of the
number of busy posts occurred before \texttt{MPI\_Waitall}.
Dividing the resulting total time by the number of operations
progressed, we measure \txprog\ to be 59.82 nanoseconds.
Less than a nanosecond of \txprog\ (due to the aforementioned
amortization) occurs in the LLP; the rest occurs in the HLP (\hlptxprog).

We include the time incurred in busy posts under \misc.
Using the tracked number of busy posts we can compute
the total time spent in busy posts during an
\texttt{MPI\_Isend}-\texttt{MPI\_Waitall} window. Dividing
this total time by the number of operations in the window
gives us an average of 3.17 nanoseconds per operation in \misc.

We use OSU Micro-Benchmark's~\cite{osu} message rate
test\footnote{We remove the send-receive sync after every window
	of posts for a clear analysis.} to measure the observed injection
overhead. By taking the inverse of the message rate, we measure
the mean injection overhead to be \textbf{263.91} nanoseconds. The
injection overhead computed with \eqref{eq:compinjection} is \textbf{264.97}
nanoseconds which is within \textbf{1\%} of the observed overhead.
\figref{fig:compinjbreakdown} shows the breakdown of the overall
injection overhead.

\begin{figure}[tbp]
	\begin{center}
		\includegraphics[width=0.49\textwidth]{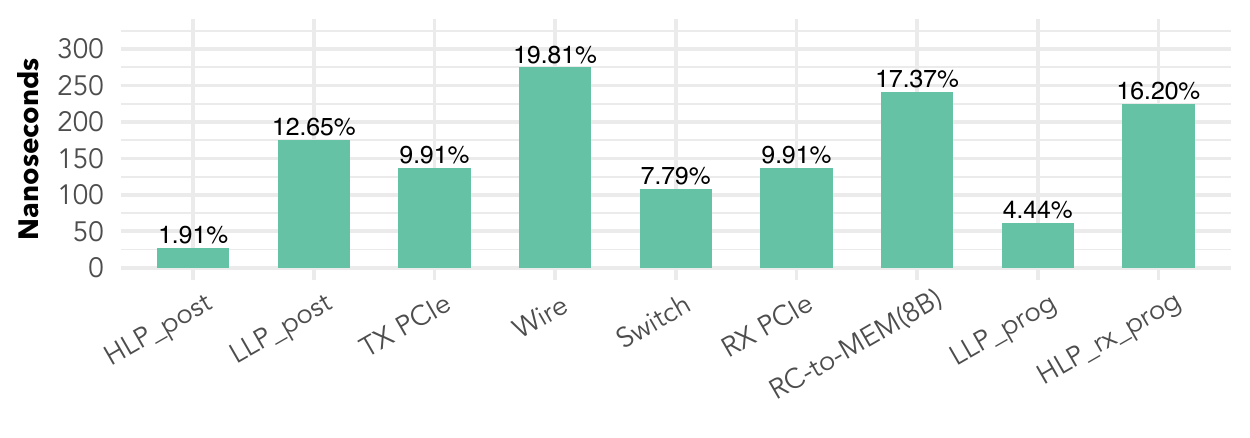}
	\end{center}
	\vspace{-1.5em}
	\caption{Breakdown of the end-to-end latency.}
		\vspace{-1.5em}
	\label{fig:fine}
\end{figure}

\minititle{End-to-end latency}. \secref{sec:latency} describes the
constituents of \latency\ with minimal software involvement.
To complete the picture, we add in the latencies of the HLP as follows.
\begin{dmath*}
\latency = \hlppost + \llppost + 2(\pcie) + \network + \rctomem[x] + \llpprog + \hlprxprog
\end{dmath*}

\hlprxprog\ refers to the overhead of progressing the reception of an
incoming message with MPI (after it has been written to memory by the RC).
We assume the initiation of the receive (such as \texttt{MPI\_Irecv})
overlaps with the rest of the constituents and, hence, do not account
for its time in the end-to-end latency. 

\hlprxprog\ is the sum of the times spent in the registered callbacks
of MPICH and UCP along with the remaining time spent in MPICH
after \texttt{ucp\_worker\_progress} returns. Note that the latter is not
the equivalent of the total time spent in MPICH for a successful
\texttt{MPI\_Wait} minus the time spent in the MPICH callback. \texttt{MPI\_Wait}
is a blocking call and incurs a portion of the 293.99 nanoseconds before even
progressing UCP. MPICH internally loops on \texttt{ucp\_worker\_progress}
until the operation is complete. Hence, we specifically measure the time
spent in MPICH after a successful \texttt{ucp\_worker\_progress} and
observe this time to be 36.89 nanoseconds. The value of
\hlprxprog\ then is 224.66 nanoseconds. Adding in the values of \llpprog,
\hlppost, and \hlprxprog\ to the modeled latency in \secref{sec:latency},
the end-to-end latency is \textbf{1387.02} nanoseconds. This is within \textbf{4\%}
of the observed latency of \textbf{1336} nanoseconds measured by OSU Micro-Benchmark's point-to-point latency test. \figref{fig:fine}
shows a detailed breakdown of this latency.

\begin{figure}[tbp]
	\begin{center}
		\includegraphics[width=0.49\textwidth]{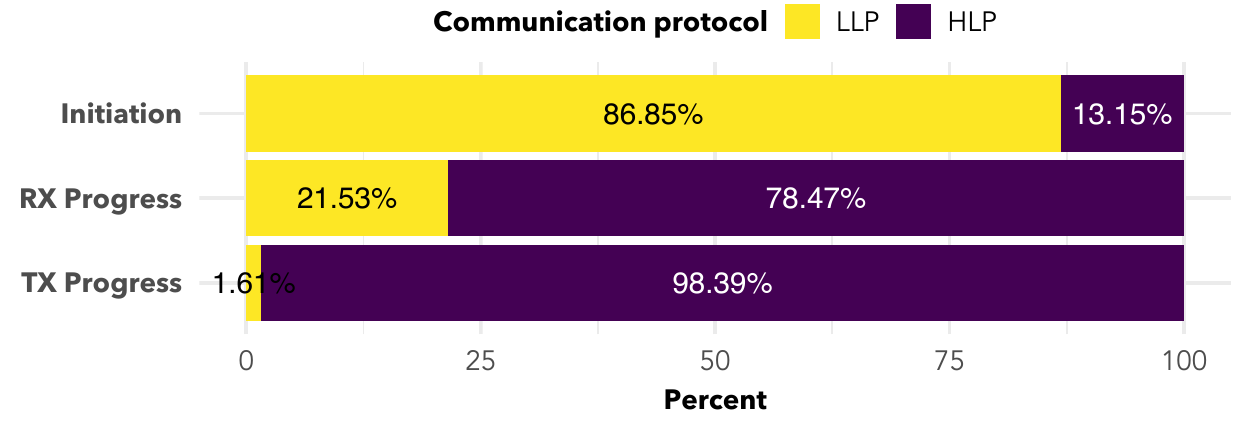}
	\end{center}
\vspace{-1em}
	\caption{Breakdown of time in HLP and LLP during the initiation and progress
		of communication.}
	\vspace{-1em}
	\label{fig:comm}
\end{figure}

\minititle{Insight 1}. \secref{sec:injection} describes that the programmer
cannot indefinitely initiate messages. Hence, the progress of a send
operation serves as a "semantic bottleneck". Once the performance
overheads imposed by this bottleneck is minimized through optimizations
like unsignaled completions, \figref{fig:compinjbreakdown} shows that \post\
dominates (more than 70\% of total) the overall injection overhead.
Within \post, the LLP dominates as seen in "Initiation" of \figref{fig:comm}.

\begin{figure}[tbp]
	\begin{center}
		\includegraphics[width=0.49\textwidth]{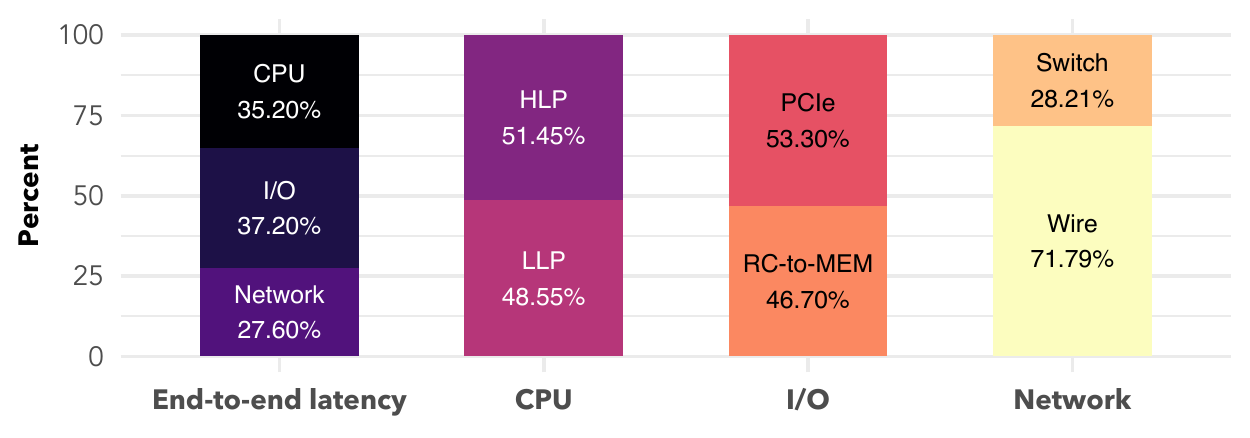}
	\end{center}
\vspace{-1em}
	\caption{High-level breakdown of the end-to-end latency.}
	\vspace{-1.25em}
	\label{fig:breakdown}
\end{figure}

\minititle{Insight 2}. \figref{fig:breakdown} presents the overall percentage
breakdown of the end-to-end latency of a small message in the three categories:
CPU, I/O, and network. The constituents of the software
and I/O categories contribute almost equally (within 4\% of each other)
to their respective total times. In the case of \network, the latency of \wire\
dominates the overall off-node time. Note that none of the three
categories dominates the overall latency. However, we observe that
the network fabric constitutes less than a third of the overall latency
while CPU and I/O components together contribute towards 72.4\%
of the latency. Hence, most of the overhead in the transmission of a
small message is incurred on the node.

\begin{figure}[tbp]
	\begin{center}
		\includegraphics[width=0.49\textwidth]{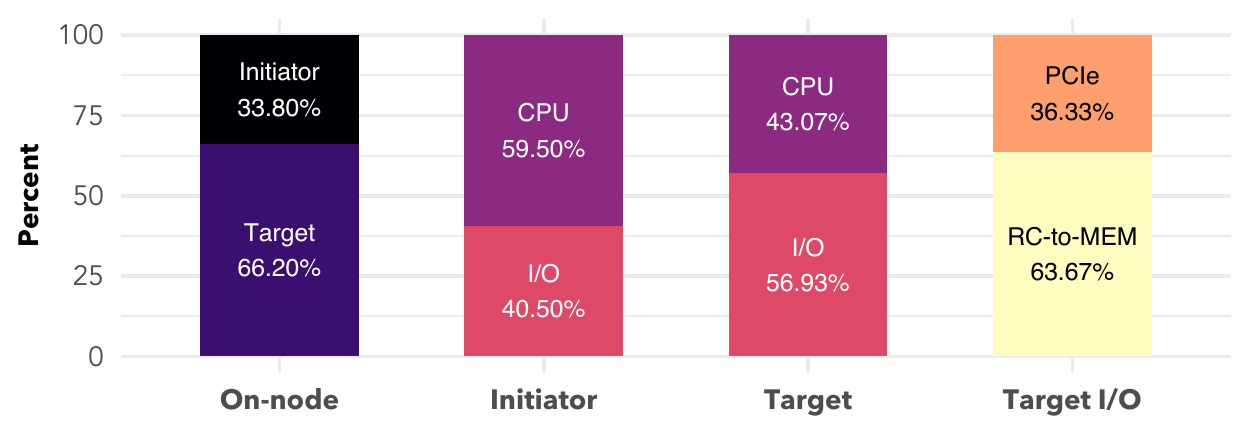}
	\end{center}
\vspace{-1em}
	\caption{Breakdown of time spent on node.}
		\vspace{-1.5em}
	\label{fig:onnode}
\end{figure}

\minititle{Insight 3}. \figref{fig:onnode} shows a high-level breakdown of the time spent on
the node during the transmission of the message. The majority of this time
occurs on the target node. Out of the time on the target
node, the majority occurs during I/O, the majority of which is comprised
by the RC writing the payload to memory. On
the contrary, software comprises the majority of the time spent on the
initiator node. This is due to the use of Programmed I/O
(see \secref{sec:background}) for short messages. Consequently, I/O
on the initiator node comprises only of a PCIe transaction unlike that on
the target node.

\minititle{Insight 4}. \figref{fig:comm} shows
that the HLP dominates the progress of both send and receive operations.
The progress of a receive operation is $4.78\times$ higher than that of a send
operation.




\section{Simulated Optimizations}
\label{sec:simulated}

\begin{figure*}
	\begin{subfigure}{.24\textwidth}
		\centering
		\includegraphics[width=\linewidth]{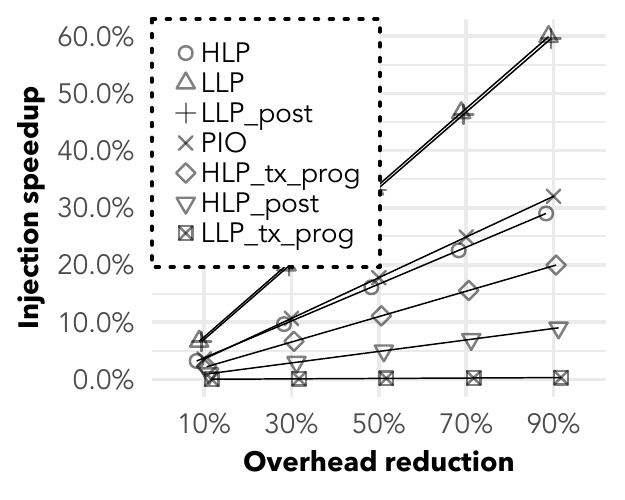}
		\caption{CPU }
		\label{fig:softwareinjwhatif}
	\end{subfigure}%
	\begin{subfigure}{.24\textwidth}
		\centering
		\includegraphics[width=\linewidth]{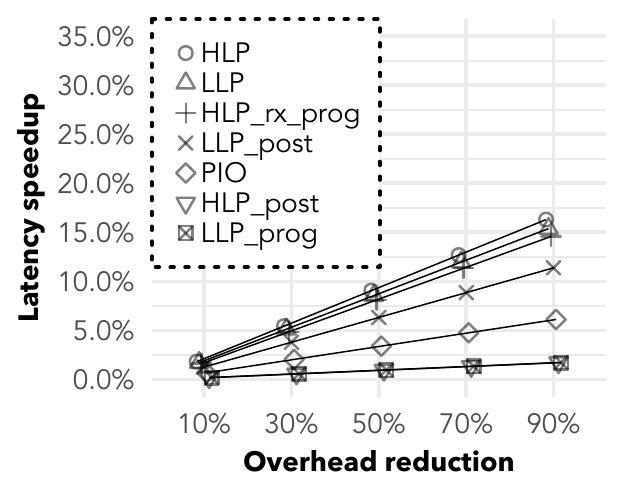}
		\caption{CPU}
		\label{fig:softwarewhatif}
	\end{subfigure}%
	\begin{subfigure}{.24\textwidth}
		\centering
		\includegraphics[width=\linewidth]{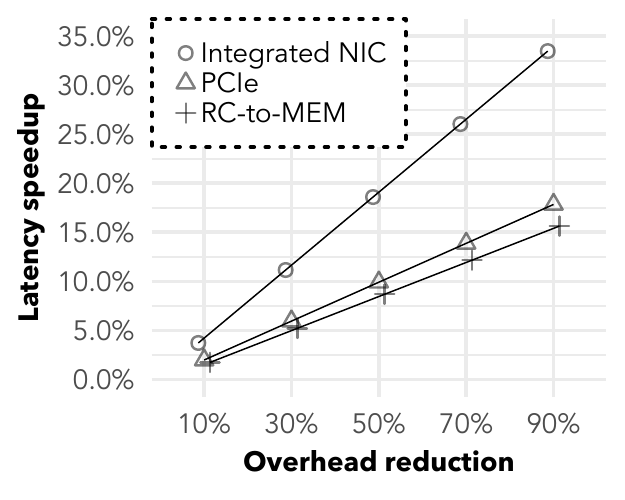}
		\caption{I/O}
		\label{fig:iowhatif}
	\end{subfigure}
	\begin{subfigure}{.24\textwidth}
	\centering
	\includegraphics[width=\linewidth]{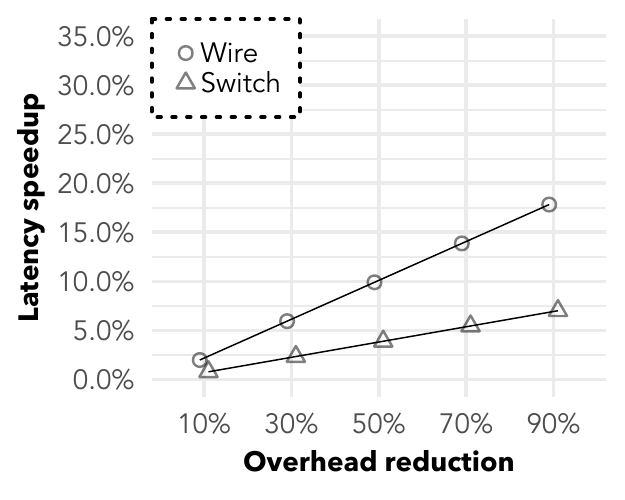}
	\caption{Network}
	\label{fig:networkwhatif}
	\end{subfigure}
	\vspace{-1em}
	\caption{Simulated speedups in overall injection (a) and end-to-end latency (b, c, d)
		by reducing overheads of CPU, I/O, and network components (note the differences
		in y-axis scales).}
	\vspace{-1em}
	\label{fig:whatif}
\end{figure*}

In this section, we use the insights gained from the breakdown of
the complete picture in \secref{sec:complete} to
study the effects of optimizing the CPU, I/O, and network fabric
components on the injection and latency of small message transfers.
In the figures that follow, we aim to answer the following question:
if we optimize component X by Y\%, what is the corresponding reduction
in injection overhead and latency? The horizontal axis of \figref{fig:whatif}
represents the degree of optimization for the component of interest.
It consists of five evenly spaced reductions in overhead, starting
from 10\% ($1.1\times$ faster) to 90\% ($10\times$ faster). The vertical axis represents
the speedup in the overall injection or end-to-end latency
as a result of reducing the component's overhead. Note that the components of our models
are not concurrent, that is, their executions do not overlap. Hence,
evaluating the impacts of reductions
in overheads on benchmarks such an MPI stencil kernel through 
a distributed system simulator (such as SimGrid~\cite{casanova:hal-01017319})
results in exactly the same linear speedups that we generate through a
manual what-if analysis in \figref{fig:whatif}.
We organize our
discussion into a set of relevant optimizations that target the different
components. For each optimization we discuss their likelihood and evaluate
their impact. We consider speedups more than 5\% to be substantial.


\vspace{-1em}
\subsection{On-node optimizations}
\label{sec:onnodeopt}

In \secref{sec:complete} we learn that most of the time in the
transmission of a small message is spent on the node. CPU
and I/O components make for the on-node time. Below
we discuss three relevant optimizations. 

\minititle{NIC integrated into a System-on-Chip (SoC).}
The idea of this optimization is that the NIC sits on the same die
as that of the processor. The deployment of such a solution
would be in the form of an SoC so that instead of interfacing with
the CPU through the PCIe subsystem, the NIC would connect to
the network-on-chip (NoC). Such a tight integration of the NIC
and the CPU would eliminate a majority of the I/O subsystem's
overhead, which accounts for the majority of the time in the
latency of a small message. While integrated NICs are not commonplace
in today's HPC systems, they are more than likely to become
ubiquitous in the future given the potential of their impact. There
have been multiple works ~\cite{binkert2006integrated, liao2009performance}
that argue for and evaluate the performance of SoC-integrated NICs
showing their benefits in terms of better performance and higher CPU
availability for all message sizes. More recently, Arm-based
supercomputers are on the rise~\cite{jackson2019evaluating} since
they allow HPC vendors to integrate their custom solutions
(such as an integrated NIC) with Arm IP on SoCs. The Tofu
interconnect D~\cite{ajima2018tofu} on Fujitsu's post-K machine is a
prominent example of this optimization. With Tofu's NIC integrated into a post-K-node, the RDMA-write latency has been improved by nearly 400 nanoseconds.

\textbf{\emph{Impact}}. "Integrated NIC" in \figref{fig:iowhatif}
shows the impact of a solution that simply brings the
NIC closer to the TX2-based SoC. While one can expect such a
solution to eliminate most of the I/O overhead, we can observe over
a 15\% improvement in overall latency even with a modest
50\% reduction in I/O time. In fact, a tightly integrated NIC
allows for opportunities to  reduce the involvement of the CPU
in the LLP's HW/SW interface and thereby increase its availability
for computational tasks. Recall that the reason for the use of
PIO for small messages is expensive PCIe round-trip latencies
with the communication-offloading approach (see \secsref{sec:intro}{sec:background}).
Since an integrated NIC would sit close to memory, round-trip
latencies performed by the hardware logic of the NIC would most
likely be faster than involving the CPU in PIO.

\textbf{Improving the initiation of a message in LLP.} This optimization
deals with how writes to device memory occur in the microarchitecture
of a processor. Ideally, writes to \texttt{aarch64}'s \emph{Device memory}~\cite{armmemory} should be as fast
as writes to its \emph{Normal memory}~\cite{armmemory}. Such an optimization is likely since
the current difference between 64-byte writes to Normal and Device
memory is more than 90\%, hinting that there exists room for optimization.
It would reduce the time spent in the PIO copy, which accounts for
more than 50\% of the time in \llppost\ (see \figref{fig:llppostbreakdown}).

\textbf{\emph{Impact}}. "PIO" in \figref{fig:softwareinjwhatif} and
\figref{fig:softwarewhatif} shows the impact of improving the
64-byte PIO copy on the overall injection and end-to-end latency,
respectively. A regular 64-byte \texttt{memcpy} on the TX2-based server
takes less than a nanosecond as expected. If we modestly project the
overhead of PIO to reduce to 15 nanoseconds (84\% reduction),
overall injection can improve by more than 25\% and end-to-end
latency can improve by more than 5\%. 

\textbf{Reducing software overheads.} This optimization
deals with software engineering targeted to reduce overheads in the
HLP. However, unlike the previous optimizations, it is unlikely that this
optimization would reduce overheads by more than 50\%. For example, the current implementation
of MPICH is highly optimized~\cite{raffenetti2017mpi}, reducing the
number of instructions by 76\% from its previous implementation for
an \texttt{MPI\_Isend}. We conjecture that software optimizations
would reduce overheads by less than 20\%.

\textbf{\emph{Impact}}. \figref{fig:softwareinjwhatif} and
\figref{fig:softwarewhatif} show the what-if analysis for the different
components in the HLP and LLP. The "HLP" and "LLP" lines in the
figures reflect the upper bound on speedups that would result from
optimizing the components that constitute the HLP and LLP, respectively.
For both injection and latency, optimizing the progress of operations in
the HLP (\hlptxprog\ and \hlprxprog) can achieve speedups close to
HLP's upper bound. Similarly, optimizing \llppost\ can achieve
speedups close to LLP's upper bound. If we consider software overheads
would be reduced at most by 20\%, the upper bounds
reflect a less than 5\% speedup in the end-to-end latency. On
the other hand, a 20\% reduction in overhead in the HLP can speedup injection
by up to 6.44\% while that in the LLP can do so by up to 13.33\%.

\vspace{-1em}
\subsection{Off-node optimizations}
\label{sec:offnodeopt}

\figref{fig:breakdown} shows that 27.6\% of the end-to-end latency
is spent on the interconnect's \wire\ and in the \switch. Our foresight
is that the reduction in off-node overheads is less than likely and
that the resulting speedups with off-node
optimizations alone would not be substantial. We explain our
foresight below.

The reduction in \wire's overhead is less than likely due to engineering
complexities at the physical layer. In fact, it is possible that the latency
will increase in future interconnects in order to accommodate for higher
throughput. The conversion between the parallel PCIe signals and the
serial signals on the interconnect's fiber transmission link occurs
through SerDes (serializer/deserializer) integrated circuits. For throughputs
higher than 100 Gb/s, the SerDes unit needs to be able to deliver higher
throughput. While higher degrees of pulse amplitude modulation (PAM)
deliver higher signal rates, they require more complex forward error
correction (FEC), which increases the latency of the transmission in some
cases by 300 nanoseconds~\cite{sun2017serdes,ghiasi2012investigation,bhoja2014fec}.

The current latency of a high-performance interconnect's switch is already
an order of magnitude lower than that of an Ethernet's switch~\cite{rumble2011s}.
New technologies like GenZ forecast their switch latencies to be 30-50
nanoseconds~\cite{genz}. However, such low latencies are
yet to be demonstrated. Only an optimistic reduction to 30 nanoseconds (72\%
overhead reduction) would correspond to a substantial speedup (5.45\%) in
end-to-end latency according to \figref{fig:networkwhatif}.

\section{Related Work}
\label{sec:related}

Prior research (described below) show the effects of optimizing
certain components on the overall communication
performance. We take an inverse approach that first
explains the observed performance and then showcases the
potential of optimizations. To the best of our
knowledge, this paper's detailed breakdown
encompassing all CPU, I/O, and network components is the
first of its kind. Additionally, such work is the first 
for an Arm-based server.

\minititle{Communication breakdown}. Papadopoulou et al.~\cite{papadopoulou2017performance}
present a detailed instruction breakdown of initiation and
progress functions between UCP and UCT to identify
engineering and abstraction overheads. Similarly,
Raffenetti et al.~\cite{raffenetti2017mpi} analyze the
overheads in the MPICH library using instruction analysis.
Both reduce the number of instructions
used in commonly used functions, resulting in higher
communication performance. However, they only focus on
one level of the stack. Our work spans both the MPICH and
UCX stacks in addition to I/O and network components.
Ajima et al.~\cite{ajima2018tofu} present a breakdown of
an RDMA-write latency on the post-K system using
simulation waveforms of hardware emulators. Our work
presents a breakdown with measured times using our described
methodology as opposed to instructions or simulations to
explain the observed communication performance. 

\minititle{Relevance of I/O}. Like us, several others also mention the bottlenecks
imposed by PCIe in datacenter networking systems.
Kalia et al.~\cite{kalia2016design} emphasize the need to
consider the low-level details and features of Verbs and the
PCIe subsystem while designing RDMA-based systems. In fact,
R. Neugebauer et al.~\cite{neugebauer2018understanding}
and Alian et al.~\cite{alian2018simulating} contribute PCIe models to
evaluate the impact of improvements to current I/O subsystems.
We quantitatively compare the I/O overheads against those
of CPU and network components. In addition to PCIe, we
profile the time spent by the RC to write to memory.
Unlike prior work, we use PCIe traces to validate software measurements.
\section{Conclusion}
\label{sec:conclusion}

Our analytical models of the injection overhead and latency of high-performance
communication on state-of-the-art components explain observed performance 
with a 5\% margin of error. The models and their resulting breakdown give the
reader insights into where, why, and how much time is spent during the
transfer of small messages. As the importance of small, fine-grained communication
is rising, we believe that such a breakdown can guide the efforts of software
developers and system architects alike to address the bottlenecks present today.
More importantly, researchers and engineers can identify bottlenecks on their
own systems using our detailed methodology described in this paper.
\begin{acks}
We thank Giri Chukkapalli and Ham Prince from Marvell Technology
Group, Yossi Itigin from Mellanox Technologies, and Pavan Balaji
from Argonne National Laboratory for their aid and support.
\end{acks}

\bibliographystyle{ACM-Reference-Format}
\bibliography{bib/refsused}

\end{document}